\documentstyle[aps,pra]{revtex}
\tightenlines
\title{On the Experimental Incompatibility Between Standard and\\
Bohmian Quantum Mechanics}
\author{M. Golshani \footnote{E-mail: golshani@ihcs.ac.ir} and O. Akhavan \footnote{E-mail:
akhavan@mehr.sharif.ac.ir}} \address{Department of Physics, Sharif
University of Technology, P.O. Box 11365--9161, Tehran,  Iran
\\Institute for Studies in Theoretical Physics and Mathematics, P.O. Box 19395--5531, Tehran, Iran}
\begin{document}
\maketitle

\begin{abstract}
Recently, three experiments have been proposed in order to show
that the standard and Bohmian quantum mechanics can have different
predictions at the individual level of particles.  However, these
thought experiments have encountered some objections.  In this
work, it is our purpose to show that our basic conclusions about
those experiments are still intact.
\end{abstract}

\pacs{ }PACS number: 03.65.Bz \maketitle

\section{Introduction}
The standard quantum mechanics (SQM) involves a set of rules that
allow physicists to evaluate statistical correlations between data
associated with the experimental procedures of preparation and
measurement, using the wave function of a system of particles. In
fact, the statistical interpretation of the wave function is in
accord with all experiments that have been performed yet.
However, Bohm {\cite{Boh}} in 1952 proposed his subquantum
realistic theory which is now often called Bohmian quantum
mechanics (BQM), with a more detailed description than SQM, so
that it deals directly with the properties of quantum objects
rather than with merely statistical results.  In other words, BQM
provides a realistic interpretation of quantum phenomena by
adding hidden variables, representing the position of particles,
to the wave function.  Hence, all particles have well-defined
positions at all times, and follow trajectories determined by the
quantum wave function $\psi$ using the guidance condition
\begin{equation}
\dot{{\bf
x}}_{i}=\frac{\hbar}{m_{i}}Im(\frac{\nabla_{i}\psi}{\psi}),
\end{equation}
which its unitary time development is governed by
Schr$\ddot{o}$dinger's equation.

Bohm's theory outlined above constitutes a consistent theory of
motion about which more details can be found in {\cite{Hol}}. By
the way, in order to ensure the compatibility of the motion of an
ensemble of particles with the results of SQM, Bohm {\cite{Boh}}
put the further constraint
\begin{equation}
P=|\psi|^{2}
\end{equation}
on the density of the probability $P$, a constraint which is
sometimes called the quantum equilibrium hypothesis (QEH).  It
should be noted that, QEH is a consistent subsidiary condition
imposed on the causal theory of motion and has no more
fundamental status than that.

However, if we deal with individual systems, we can hope that the
determination of one of the initial positions of the system,
using an ingenious way, provides grounds of differentiation
between SQM and BQM, although because of QEH, it is evident that
they certainly give the same statistical prediction for an
ensemble of the individual particles.  Concerning this point,
recently Ghose {\cite{Gho}} and these authors [4-6] proposed some
experiments which could differentiate SQM from BQM. But,
Marchildon {\cite{Mar}} accompanied by Struyve and De Baere
{\cite{Str}} have claimed that the proposed experiments cannot
provide different predictions for SQM and BQM.  Although Ghose
[9,10] and these authors [5,6] believe that Marchildon's
objections do not change our main results, Marchildon still holds
that our constraint on $y_{0}$ in [5,6] is doubtful {\cite{Marc}}.

In this work, we have presented a concise review of our three
experiments and a more detailed discussion on them in order to
make firm our previous basic conclusions, particularly against the
objections raised in [7,8].

\section{An outline along with a more detailed discussion about the three
suggested experiments} In order to have a more complete
examination on the three suggested experiments in [3-6], we have
presented a summary of each experiment separately, as well as our
reasons against the objections raised in [7,8].
\subsection{A two-particle double-slit experiment with an entangled
wave function} Consider a double-slit screen with two identical
slits $A$ and $B$ with the width $2\sigma_{0}$ and centers
located at $(0,\pm Y)$, in the $x-y$ plane.  Instead of the usual
one particle emitting source, consider a special point source
emitting pairs of identical non-relativistic entangled particles,
and which is located very far from the two-slit screen. The
entanglement of particles 1 and 2 is described by the following
conditions:
\begin{eqnarray}
&&y_{1}+y_{2}=0,\cr&&
p_{1y}-p_{2y}=0.
\end{eqnarray}
The general wave function of this two-particle system, after
diffraction from the two slits, is given by [3,5]
\begin{eqnarray}
\psi(x_{1},y_{1};x_{2},y_{2};t) =
N[\psi_{A}(x_{1},y_{1},t)\psi_{B}(x_{2},y_{2},t)\pm\psi_{A}(x_{2},y_{2},t)\psi_{B}(x_{1},y_{1},t)].
\end{eqnarray}
To have a complete discussion, we assume that the two slits
produce the Gaussian wave packets along the $y$-direction at
$t=0$, that is,
\begin{equation}
\psi_{A,B}(x,y,t)=(2\pi \sigma_{t}^{2})^{-1/4}e^{[-(\pm
y-Y-u_{y}t)^{2}/4\sigma_{0}\sigma_{t}+ik_{y}(\pm
y-Y-u_{y}t/2)]}e^{i(k_{x}x-E_{x}t/\hbar)},
\end{equation}
where
\begin{equation}
\sigma_{t}=\sigma_{0}(1+\frac{i\hbar t}{2m\sigma_{0}^{2}}),
\end{equation}
and
\begin{equation}
u_{y}=\frac{\hbar k_{y}}{m}, \hspace{5mm}
E_{x}=\frac{1}{2}mu_{x}^{2}.
\end{equation}
Based on SQM, it is well known that the probability of
simultaneous detection of the pair of particles, at arbitrary
points $y_{1}=Q_{1}$ and $y_{2}=Q_{2}$ on the screen, is
\begin{equation}
P_{12}(Q_{1},Q_{2},t)=\int_{Q_{1}}^{{Q_{1}}+\triangle}\int_{Q_{2}}^{Q_{2}+\triangle}dy_{1}dy_{2}|\psi(y_{1},y_{2},t)|^{2},
\end{equation}
where $\triangle$ represents the width of a particle detector on
the screen.

On the other hand, using BQM, we obtained the equation of motion
for the $y$-coordinate of the center of mass of the two particles
in ref. {\cite{Gols}} as
\begin{equation}
y(t)=y_{0}\sqrt{1+(\hbar t/2m\sigma_{0}^{2})^{2}},
\end{equation}
where $y=(y_{1}+y_{2})/2$, and $y_{0}$ is the vertical coordinate
of the center of mass at $t=0$.  It is clear that, if the
condition $y_{0}=0$ is considered, then the center of mass will
remain on the $x$-axis for all times.  Thus, based on BQM, each
entangled pair of particles will be always detected symmetrically
with respect to the $x$-axis.  However, SQM predicts that the
probability of asymmetrical detection of the pairs of particles
can be different from zero, at variance with BQM's symmetrical
prediction. Furthermore, according to SQM's prediction, the
probability of finding two particles at one side of the $x$-axis
can be non-zero, while it is shown that BQM forbids such events,
provided that $y_{0}=0$.

Here, somebody may feel reluctant to use the $y_{0}=0$ constraint
(for example, [7,8]).  In fact, one may argue that $y_{0}$ must be
distributed according to $|\psi|^{2}$, because of QEH.  We agree
that the properties of the individual particles, that is,
$y_{1}$, $y_{2}$, $p_{1y}$ and $p_{2y}$, are undetermined based on
QEH, but it should be noted that their joint properties are
completely defined, as shown on page 77 of ref. {\cite{Bou}} and
in the following.  In fact, although the operators $\widehat{y}$
and $\widehat{p}_{y}$ do not commute for each particle,
\begin{equation}
[\widehat{y}_{i},\widehat{p}_{iy}]=i\hbar,
\end{equation}
but the operators for $(\widehat{y}_{1}+\widehat{y}_{2})$ and
$(\widehat{p}_{1y}-\widehat{p}_{2y})$ do commute, i.e.,
\begin{equation}
[(\widehat{y}_{1}+\widehat{y}_{2}),(\widehat{p}_{1y}-\widehat{p}_{2y})]=0.
\end{equation}
Therefore, for the entangled wave function, the joint properties
$(y_{1}+y_{2})$ and $(p_{1y}-p_{2y})$ could both be determined
with an arbitrary accuracy. In addition, using this conclusion,
the objection of Struyve and De Baere {\cite{Str}} about the
non-ergodic property of BQM, which is one of Ghose's arguments
about the incompatibility of SQM and the conventional de
Brogli-Bohm theory [10,11], cannot be sustained.

However, we also studied a case in which we can have $\triangle
y_{0}\neq 0$ and $\langle y_0\rangle=0$.  It is shown that, for
obtaining symmetrical detection with a reasonable approximation
it is enough to require the following constraint {\cite{Gols}}
\begin{equation}
0\leq y_0(t) \ll\sigma_0,
\end{equation}
where $y_0(t)$ shows variation of $y_0$ with time.  Once again,
due to the entanglement of the particles, it is a possible
constraint, although each particle has its quantum equilibrium
distribution, that is,
\begin{equation}
(\triangle y_{1})_{t=0}=(\triangle y_{2})_{t=0}\sim\sigma_0.
\end{equation}
In fact, contrary to the single-particle two-slit experiment which
always requires the constraint $\triangle y_0\sim\sigma_0$,
because of the uncertainty on both particle's position and
momentum, the case of two-entangled particle double-slit
experiment has the advantage of $y_0$ determination with the
required accuracy and so, for example, we do not need to take very
small slits in order to make sure that the particles depart from
the $x$-axis symmetrically, as was offered by Struyve and De Baere
{\cite{Str}}.

On the other hand, concerning the $y_0$ determination, Marchildon
{\cite{Marc}} believes that once particles have gone through the
slits, much of their memory of coming from the source is erased.
He has argued that, if the wave packets coming out of the slits
are as in eq. (5) at $t=0$, then the $y$-coordinates are spread
independently, with the standard deviations of the order of
$\sigma_0$, and the center of mass coordinate $y$ has a similar
spread.  To answer this kind of criticism, we should investigate
the issue of entanglement of particles, using both SQM and BQM.

Using SQM, one can think of the entanglement of the two particles
at the source, in the form of the two following alternatives:\\
${\it 1}$. The entanglement constraint
$(y_{1}+y_{2})_{t\rightarrow -\infty}=0$ at the source, is erased
during the diffraction of the pair at the slits at $t=0$.  Thus,
the results of the joint probability in eq. (8) bcomes
inconsistent with BQM's symmetrical prediction which still
maintains the entanglement property of the two particles.  The
validity of the entanglement of the two particles in BQM after
the diffraction, has been further discussed in the following.\\
${\it 2}$. SQM agrees that the entanglement property is
maintained at all times, i.e. $y_{1}(t)+y_{2}(t)=0$, because
there is no interaction between the pair and the double-slit
screen and, all conditions are identical for the two particles.
Then, the joint probability density is given by
\begin{equation}
P_{12}(y_{1}(t),-y_{1}(t),t)=|\psi(y_{1}(t),-y_{1}(t),t)|^{2}.
\end{equation}
But, there is a problem here.  It is clear that, the joint
probability density
\begin{equation}
P_{12}(y_{1}(t),y_{2}(t),t)=|\psi(y_{1}(t),y_{2}(t),t)|^{2}\hspace{1cm}
if:\hspace{4mm} y_{1}(t)+y_{2}(t)\neq 0,
\end{equation}
has a non-zero value.  In fact, $P_{12}(y_{1}(t),-y_{1}(t),t)$
cannot be equal to 1 for every entangled pair, as was mentioned
by Struyve and De Baere {\cite{Str}}, if one uses SQM.  However,
these authors use SQM to refute our statements about BQM.\\
Thus, we should accept that either the asymmetrical detection of
the two particles with erased entanglement is a right prediction
or that SQM is an incomplete theory.  But, accepting the first
alternative means inconsistency between SQM and BQM, as we have
shown in the following.

It is well known in BQM that, the wave function $\psi$ plays two
conceptually different roles:\\
{\it 1}. As determining the influence of the environment on the
particle via the guidance condition (1).\\
{\it 2}. As determining the probability density $P=|\psi|^{2}$,
i.e. QEH.\\
But, the primary conceptual role for $\psi$ in Bohm's theory is
the first role.  In fact, in BQM, probability only enters as a
subsidiary condition on a causal theory of the motion of
individual particles in order to make statistical prediction
equivalent with SQM, i.e., the statistical meaning of the wave
function is a secondary property.  In our experiment too, BQM
requires that the entanglement of the two particles must be
considered in their tracks, so that the quantum distribution of
the pairs on the screen is formed according to the equation of
motion
\begin{equation}
y_{1}(t)+y_{2}(t)=2y_{0}\sqrt{1+(\hbar t/2m\sigma_{0}^{2}})^{2},
\end{equation}
which is weighted by the probability density $P=|\psi|^{2}$. By
considering the equation of motion for the center of mass
coordinate $y$, we shall have the time development of the
entanglement property of the two particles which originated from
the source at $t\rightarrow  -\infty$.  Hence, in BQM, rather than
SQM, the entanglement of the two particles is not forgotten.  By
the way, it is shown that the final interference pattern is the
same as the one predicted by SQM [3,5], which shows the
consistency of the result obtained for the ensemble of particles
with QEH. It is worthy to note that, in Bohm's theory, underlying
quantum mechanics is a causal theory of the motion of waves and
particles, which is consistent with a probabilistic
interpretation, but does not require it. Therefore, contrary to
the case of SQM, in BQM, which is a causal theory, the
entanglement is not erased and in consequence, SQM's probabilistic
prediction must differ from BQM's deterministic prediction, at
the individual level, although the wave function used for the two
theories as well as their final predicted interference patterns
are the same.

Although we have studied the important objections to GGA
experiment [3,5], it is useful to have a more detailed discussion
about some of the remaining objections.  Marchildon {\cite{Mar}}
as well as Struyve and De baere {\cite{Str}} claim that for $\hbar
t/2m\sigma_{0}^{2}\ll 1$, SQM also predicts symmetrical detection
as BQM does. Although we agree that under this condition the
predictions of the two theories approaches the same results, we
still hold that SQM's and BQM's results for this experiment are
really different, because Gaussian slits with the width
$2\sigma_0$ and the distance $2Y\sim\sigma_0$ produces a small
but finite overlapping of the two wave functions in the domain
$|y|<Y$  on the screen which prevents symmetrical prediction,
using SQM.  In addition, to make the subject clearer, we can
consider either $\sigma_{0}\longrightarrow 0$ and
$m\longrightarrow\infty$ or any other desired variations that
could produce
\begin{equation}
\frac{\hbar t}{2m\sigma_{0}^{2}}\sim 1.
\end{equation}
This condition yields $y\sim y_0$ due to eq. (9).  By considering
$Y\sim \sigma_0$, eq. (17) means that there is a considerable
overlap of the two wave packets emerging from the two slits on the
screen and, in consequence, SQM does not predict symmetrical
detection.  However, BQM can predict acceptable symmetrical
detection if the constraint
\begin{equation}
0\leq\triangle y_0\ll \sigma_0
\end{equation}
is applied, which is feasible for the two entangled particles.
Thus, our previous basic conclusion in refs. [5,6] is still
unchanged, when this condition holds.

On the other hand, Struyve and De Baere {\cite{Str}} stated that,
if, e.g., $\sigma_0$ is considered very small so that $\hbar
t/2m\sigma_{0}^{2}\gg 1$, then an asymmetrical detection on the
screen is predicted by BQM, although $y_0$ was considered very
small and we assured that the particles depart the $x$-axis
symmetrically.  Clearly, we agree with this special case. But, we
have shown that for the two entangled particles, $y_0$ is
adjustable at the source with a desired precision and we do not
need to take $\sigma_0$ very small to make sure of symmetrical
departure of the two particles from the $x$-axis. In addition,
for the case of a double-slit experiment with two unentangled
particles, we have shown that, even this special case can lead to
a different prediction between SQM and BQM, using selective
detection [4,5]. We have considered the discussion of this
experiment in the next subsection.

\subsection{A two-particle double-slit experiment with an unentangled
wave function} In this experiment [4,5], we have considered a
special form of the last experimental set-up, and our source is
substituted with another one that emits two unentangled identical
particles. Hence, the wave function of this two-particle system
is given by
\begin{equation}
\psi(x_{1},y_{1};x_{2},y_{2};t)=\overline{N}[\psi_{A}(x_{1},y_{1},t)+\psi_{B}(x_{1},y_{1},t)][\psi_{A}(x_{2},y_{2},t)+\psi_{B}(x_{2},y_{2},t)],
\end{equation}
where $\psi_A$ and $\psi_B$ can be considered as the Gaussian
wave packets (5).  SQM's probabilistic prediction about the joint
detection of the two particles on the screen is the same as the
one noted for the last experiment, i.e. equation (8).

On the other hand, according to BQM, the velocity of the center of
mass of the two particles in the $y$-direction is [4,5]
\begin{equation}
\dot{y}=\frac{(\hbar/2m\sigma_{0}^{2})^{2}yt}{1+(\hbar/2m\sigma_{0}^{2})^{2}t^{2}}+\overline{N}\frac{\hbar}{2m}Im\{\frac{1}{\psi}(\frac{Y+u_yt}{\sigma_0\sigma_t}+2ik_y)(\psi_{A_1}\psi_{A_2}-\psi_{B_1}\psi_{B_2})\},
\end{equation}
so that the equation of motion for the $y$-coordinate of the
center of mass can be written as
\begin{equation} y(t)\simeq y_{0}\sqrt{1+(\hbar
t/2m\sigma_{0}^{2})^{2}},
\end{equation}
if the conditions $Y\ll\sigma_{0}$ and $k_{y}\simeq 0$ are
satisfied.  Furthermore, the guidance condition (1) yields
\begin{eqnarray}
&&\dot{y}_{1}(x_{1},y_{1};t)=-\dot{y}_{1}(x_{1},-y_{1};t),\cr&&
\dot{y}_{2}(x_{2},y_{2};t)=-\dot{y}_{2}(x_{2},-y_{2};t),
\end{eqnarray}
which imply the $y$-component of the velocity of each particle
must vanish on the $x$-axis, independent of the other particle's
position. If the case $y_0=0$ were possible, then one could have
reconsidered the last discussion on SQM's probable asymmetrical
prediction and BQM's symmetrical one.  But, as it is well known,
we agree that for this two-unentangled particle experiment, we
have
\begin{equation}
\triangle y_0\sim\sigma_0,
\end{equation}
according to QEH.  However, we still can obtain different
predictions for the two theories.  To show this, consider the case
of $\triangle y_0\sim\sigma_0$ and $\langle y_0\rangle=0$. We
assume that to obtain symmetrical detection around the $x$-axis
with reasonable approximation, it is enough that the center of
mass variation be smaller than the distance between any two
neighboring maxima on the screen. Then, we have {\cite{Gols}}
\begin{equation}
\triangle y_0\ll\frac{\pi\hbar t}{Ym},
\end{equation}
which yields
\begin{equation}
Y\ll2\pi\sigma_0,
\end{equation}
where in (24) the condition $\hbar t/2m\sigma_{0}^{2}\sim 1$ is
assumed, so that we have $y\sim y_0$.  Thus, for obtaining
symmetrical detection in BQM, we should guarantee that the
conditions $Y\ll 2\pi\sigma_0$ and $\hbar t/2m\sigma_{0}^{2}\sim
1$ be satisfied for this experiment. It should be noted that,
under these conditions, the two wave packets are overlapped on the
screen in an interval of the order of $\sigma_0$.  In this
interval neither BQM nor SQM predict symmetrical detection around
the $x$-axis.  In fact, the symmetrical detection predicted by BQM
happens at far (relative to $\sigma_0$) from the $x$-axis on the
screen.  In other words, save the central peak, which does not
show symmetry with respect to the $x$-axis, other less prominent
maxima appear at the locations
\begin{eqnarray}
&&y_{n_{+}}\simeq n_+\frac{\pi\hbar t}{Ym}\pm\triangle y,\cr&&
y_{n_{-}}\simeq -n_-\frac{\pi\hbar t}{Ym}\pm\triangle y,
\end{eqnarray}
where $y_{n_{\pm}}$ refer to $y$-component of the maxima above or
below the $x$-axis on the screen, respectively.  In addition,
$n_{\pm}$ represent positive integer numbers.  BQM's symmetrical
prediction puts the following constraint:
\begin{equation}
n_+=n_-.
\end{equation}
However, SQM's probabilistic prediction does not require this
constraint.  Note that, under the condition $Y\ll\sigma_0$, we
have
\begin{equation}
\frac{\pi\hbar t}{Ym}\gg\triangle y,
\end{equation}
because for the condition $\hbar t/2m\sigma_{0}^{2}\sim 1$, we
have $\triangle y\sim\triangle y_0\sim\sigma_0$.  That is, in BQM,
we have symmetrical peaks up to $\sigma_0$.  It is clear that,
the wave function (19) predicts the same interference patterns
for the ensemble of particles for the two theories, according to
QEH.  By the way, the same discussion on the determination of the
conditions for the case of $\triangle y_0\sim\sigma_0$ can be
applied as well to the case of a double-slit experiment with two
entangled particles.

Furthermore, for the case of $\hbar t/2m\sigma_{0}^{2}\gg 1$,
using eqs. (21) and (22) and selective detection of the two
particles, which requires registration of those two particles
that are detected at the two sides of the $x$-axis simultaneously
and omission of the others, BQM can predict a rather empty
interval with low intensity of particles that has a length
\begin{equation}
L\simeq 2\langle y\rangle\simeq\frac{\hbar
t}{m\sigma_{0}^{2}}\langle y_{0}\rangle,
\end{equation}
if the constraint $\triangle y\ll L$ is satisfied.  The last
constraint at $\hbar t/2m\sigma_{0}^{2}\gg 1$ condition,
corresponds to $\triangle y_{0}\ll\langle y_{0}\rangle$.
Therefore, it is shown that, according to BQM and under the
conditions
\begin{eqnarray}
&&\frac{\hbar t}{2m\sigma_{0}^{2}}\gg 1, \cr&&
Y\ll\sigma_0\ll\langle y_0\rangle,
\end{eqnarray}
a considerable position change in the $y$-coordinate of the source
produces a region with very low intensity on the screen which is
not predicted by SQM.  Therefore, we have disagreement between
the two theories' predictions even if $\triangle y_0\sim\sigma_0$
is considered as a constraint.

However, since the two particles in this experiment are emitted
in an unentangled state, the results obtained maybe seem
unbelievable.  In this regard, Struyve {\cite{Stru}} based on our
aforementioned factorizable wave function (19), believes that the
two independent particles of this experiment cannot produce
different predictions for SQM and BQM.  He argues that
{\cite{Stru}}, the results of the experiment will not be altered
if we emit the two particles simultaneously or emit only one
particle at a time, because the two particles are totally
independent.

Although we agree along with him that the results of this
experiment are rather strange, we believe that discussion of this
experiment can help to make the SQM and BQM disagreement more
exciting.  Hence, in the following, we substantiate our previous
arguments about this experiment.

At first, we examine the applied condition $Y\ll\sigma_0$, in a
double-slit experiment with two unentangled particles.  One may
argue that this condition is meaningless, because based on the
specifications of the set-up, Y represents the distance between
the center of each slit to the $x$-axis, and therefore, the
minimum value of $Y$ approaches $\epsilon +\sigma_0$, where
$\epsilon$ is considered very small and represents the length of
plane that separates the two slits.  But, this objection can be
answered by considering the overlapping of each particle's two
Gaussian wave functions which are generated at the two near
slits.  The overlapping causes that the peak of each Gaussian wave
approaches more and more the $x$-axis.  In addition, under this
condition, the Gaussian wave functions lose their symmetrical form
at each slit.  Our argument becomes more clearer when we consider
$\epsilon=0$ as a limiting case, i.e, we have only one slit.  In
this limiting case, it is clear that $Y=0$.  Therefore, when the
two slits are very near together, the peak of Gaussian wave
functions, i.e. $Y$, come very near to the $x$-axis and the
condition $Y\ll\sigma_0$ is completely right.

Another problem can be raised when one thinks of the two
independent particles, as mentioned by Struyve too {\cite{Stru}}.
To handle to this problem, let us reconsider the second term in
eq. (20) particularly the coefficient
$(\psi_{A_1}\psi_{A_2}-\psi_{B_1}\psi_{B_2})$.  Using the
Gaussian wave (5) and the condition $k_y\simeq 0$, the latter
coefficient can be written in the form
\begin{eqnarray}
\psi_{A_1}\psi_{A_2}-\psi_{B_1}\psi_{B_2}=&&(2\pi\sigma_t)^{-1/4}e^{2i(k_xx-E_xt/\hbar)}e^{-(y_{1}^{2}+y_{2}^{2})/4\sigma_0\sigma_t}e^{-(Y+u_yt)^2/2\sigma_0\sigma_t}\nonumber\\
&\times&[e^{(y_1+y_2)(Y+u_yt)/2\sigma_0\sigma_t}-e^{-(y_1+y_2)(Y+u_yt)/2\sigma_0\sigma_t}].
\end{eqnarray}
If we require that $y_1(t)+y_2(t)=0$, i.e., the $y$-component of
the two particles are entangled, then we would obtain the
equation of motion (9), as expected.  But our two particles in
this experiment are initially unentangled and it is not necessary
to have $y_1(t)+y_2(t)=0$.  Instead, we can have another
selection on the two-slit set-up.  In fact, if we apply the
condition $Y\ll\sigma_0$, again the behavior of the equation of
motion of the two particles in the y-direction is similar to the
motion of the two entangled particles, while the two particles
were unentangled.  Hence, we can state that the classical
interaction of the wave function of the two unentangled particles
with the two-slit plane barrier for the condition $Y\ll\sigma_0$,
results in a wave function which now guides the $y$-component of
the center of mass of the two apparently unentangled particles in
the same way as the case of two entangled particles with the
initial condition $-\sigma_0\leq(y_1+y_2)_{t=0}\leq\sigma_0$, for
those pairs of particles that pass through the two slits.  Thus,
we have shown that the results obtained in the two-slit
experiment using two synchronized identical particles and the
selective detection, are completely different from the ones
obtained in a single-particle double-slit experiment, contrary to
Struyve's belief {\cite{Stru}}.  In fact, the motion of either
particle is now dependent on its own location and the location of
the other particle, although the apparent form of the wave
function of the system can be efficiently represented by the use
of the unentangled form in (19), which is only useful at the
ensemble level of particles. Therefore, our previous basic results
about this experiment still remain intact.

\subsection{An experiment with two double slits and two entangled
particles} The set-up of this experiment {\cite{Golsh}} consists
of two double-slit screens where the slits $A$ and $B$ are on the
right screen and $A^{'}$ and $B^{'}$ on the left screen, with
their centers located at the points $(\pm d,\pm Y)$ in a
two-dimensional coordinate system. A special source which emits
pairs of identical non-relativistic entangled particles is placed
at the origin of the coordinates.

The entanglement property of the two particles is expressed by
\begin{eqnarray}
&&x_{1}+x_{2}=y_{1}+y_{2}=0,\cr&& p_{1x}-p_{2x}=p_{1y}-p_{2y}=0.
\end{eqnarray}
As we mentioned previously in subsection II. A, since we have
\begin{equation}
[(\widehat{x}_{1}+\widehat{x}_{2}),(\widehat{p}_{1x}-\widehat{p}_{2x})]=[(\widehat{y}_{1}+\widehat{y}_{2}),(\widehat{p}_{1y}-\widehat{p}_{2y})]=0,
\end{equation}
the joint properties $(x_{1}+x_{2})$ and $(p_{1x}-p_{2x})$ as
well as $(y_{1}+y_{2})$ and $(p_{1y}-p_{2y})$ can both be
determined with the desired accuracy.  Thus, for example,
determination of $y_0=\frac{1}{2}(y_1+y_2)_{t=0}$ is
theoretically possible.  Since the source used in this experiment
is the same as the one used in the EPR experiment {\cite{Ein}}, it
seems that performing this experiment can be considered more
feasible than those of the last two aforementioned experiments.
In addition, such sources can be utilized in some interesting
processes of the quantum information theory {\cite{Bou}} such as
quantum dense coding and quantum teleportation protocols
{\cite{Akhava}}.

The general form of the wave function for this system can be
written as
\begin{eqnarray}
\psi(x_{1},y_{1};x_{2},y_{2};t) =&\widetilde{N}&[\psi_{A}(x_{1},y_{1},t)\psi_{B^{'}}(x_{2},y_{2},t)\pm\psi_{A}(x_{2},y_{2},t)\psi_{B^{'}}(x_{1},y_{1},t)\nonumber\\
 &+&\psi_{B} (x_{1},y_{1},t)\psi_{A^{'}}(x_{2},y_{2},t)\pm\psi_{B}
(x_{2},y_{2},t)\psi_{A^{'}} (x_{1},y_{1},t)].
\end{eqnarray}
Using BQM, it is straightforward to show that eq. (9) again
determines the motion of the center of mass of the $y$-coordinate
of the two entangled particles.

Once again, we can have a similar discussion on SQM's
probabilistic asymmetrical prediction against BQM's deterministic
and symmetrical prediction, with the $y_0=0$ condition.  In
addition, for the conditions
\begin{eqnarray}
&&\frac{\hbar t}{2m\sigma_{0}^{2}}\sim 1,\cr&&
Y\sim\sigma_0,\cr&&
0\leq\triangle y_0\ll\sigma_0,
\end{eqnarray}
one can again obtain a reasonable symmetrical prediction, based
on BQM.  It is worthy to note that, the first two conditions
provide a considerable overlap of the wave packets on the screen,
so that SQM cannot predict a symmetrical detection.  Therefore,
our all previous conclusions about this experiment are unchanged
as well.

\section{Conclusion}
Three recent proposed experiments have been studied in some
details.  It is shown that, the objections raised by Marchildon
[7,9] as well as Struyve and De Baere [8,13] are not justified
and that the basic conclusion of the experiments, that is, the
existence of the incompatibility between SQM's probabilistic
prediction and BQM's deterministic prediction, still stands out.
In fact, they tried to show equivalence of the two theories at
the individual level of the suggested experiments, by applying
SQM's rules to both theories.  For the present experimental
ability the only difficulty seems to be related to the special
properties of the sources which produces the pairs of entangled
fermionic particles, such as electrons, neutrons or entangled
bosonic particles. However, it seems that such experiments may be
done using photons at Pavia, as Ghose promised in [10,11].
Furthermore, a new and more feasible experimental set-up to
distinguish between SQM and BQM has been suggested elsewhere
{\cite{Akhavan}} which will make the discussion on the
disagreement between the two theories more serious and more
interesting than what has been done so far.

\begin{acknowledgments}
The authors would like to thank P. Ghose, E.R. Floyd, L.
Marchildon, W. Struyve and W. De Baere as well as anonymous
referees of J. Phys. A for their valuable comments on our earlier
works.
\end{acknowledgments}

\end{document}